# Low-Damage High-Throughput Grazing-Angle Sputter Deposition on Graphene

C.-T. Chen[1,a], E. A. Casu[1,2,b], M. Gajek[1] and S. Raoux[1]

[1] IBM Thomas J. Watson Research Center, Yorktown Heights, New York 10598, USA

[2] Politecnico di Torino, Turin 10129, Italy

Despite the prevalence of sputter deposition in the microelectronics industry, it has seen very limited applications for graphene electronics. In this letter, we report systematic investigation of the sputtering induced damages in graphene and identify the energetic sputtering gas neutrals as the primary cause of graphene disorder. We further demonstrate a grazing-incidence sputtering configuration that strongly suppresses fast neutral bombardment and retains graphene structure integrity, creating considerably lower damage than electron-beam evaporation. Such sputtering technique yields fully-covered, smooth thin dielectric films, highlighting its potential for contact metals, gate oxides, and tunnel barriers fabrication in graphene device applications.

[a] Email: cchen3@us.ibm.com
[b] Present address: NANOLAB, École Polytechnique Fédérale de Lausanne, Lausanne 1015, Switzerland

Graphene exhibits excellent physical properties,[1] such as exceptional intrinsic mobility, high rigidity, fast photo-response, small spin-orbital coupling and minute nuclear spin fluctuations, which makes it a promising material for hosting future-generation nano-devices.[2-5] To turn the prospect into reality, a reliable low-damage, high-throughput method for depositing a wide range of metals and dielectrics on graphene is crucial. Currently semiconductor industry relies on sputter deposition in large-scale production of integrated circuits, memory, and other devices.[6] Sputtering can achieve high deposition rate regardless of the melting temperatures of the target materials. Moreover, sputtering generally preserves film stoichiometry when processing complex compound materials,[7] significantly expanding the scope of the technology across various industries. However, the application of sputter deposition on graphene has hitherto been limited despite the numerous advantages. Researchers mostly depend on electron-beam (e-beam) evaporation and atomic layer deposition (ALD) for metals or dielectrics deposition, putting considerable constraints on the material choice of graphene devices. The reason for the scarcity lies in the extensive damage sustained by graphene in conventional sputtering.[8,9] Previous reports[8,9] show that Raman spectra exhibit substantial spectral weight in the defect mediated D-mode of graphene after sputtering, with the exception of Ref. 10. Often the graphene G and 2D peaks also lose their characteristic shapes, indicative of amorphization of $sp^2$ C–C bonds.[11] In this paper, we systematically investigate the leading cause of structural damages during sputtering. We demonstrate that by placing of the substrates in the *grazing-angle* sample-target configuration, one can significantly reduce the impact to graphene while maintaining high deposition rates, promoting sputter deposition as a viable technology in graphene device fabrication.

We used flake graphene exfoliated from Kish graphite and deposited on Si substrates as our samples. Thin films on graphene were deposited in two magnetron sputtering systems. The majority of the parameter-tuning experiments were done in a custom-made system using small magnetron guns holding 1" diameter targets. Effects of the incidence angle, discharge power, sputtering gas species, gas pressure, deposition rate, and deposition time were all analyzed. The optimal condition was then implemented in a commercial sputtering system using sputter guns with 2" diameter targets for high-throughput and best-quality films. Both systems routinely achieved base pressure of $\leq 10^{-7}$ Torr. Most samples were grown using DC sputtering in 2 – 4 mTorr Ar with the substrate-target distance maintained at 20 – 23 cm. Structural damages in various sputtering configurations were characterized by Raman spectroscopy under 532 nm laser illuminations. Spectral weights of graphene modes were obtained by fitting to Lorentzian function. Since symmetry dictates that the Raman D mode appears only when structural defects exist, we therefore use the spectral weight ratio ($I_D/I_G$) as a quantitative measure of sputtering damage. From $I_D/I_G$, one can then derive the average domain size ($L_a$) of the resulting graphene nanocrystallites using the following formula, $L_a$ (nm) = $(2.4 \times 10^{-10}) \lambda^4_{laser}$ (nm)$(I_D/I_G)^{-1}$.[12]

To elucidate the correlation between the incidence angle and the degree of disorder induced by sputtering in graphene, we focus on two distinct substrate-target alignment configurations: *normal incidence* vs. *tangential-incidence* as the limiting cases of grazing-angle sputtering. In conventional normal-incidence sputtering, the substrate faces the target directly such that the sputtered species arrive at the substrate along its normal direction. In tangential-incidence sputtering, the tangent of the substrate surface lies in the substrate-target direction (see schematics in Fig. 1(a)), which ensures that all high-energy particles arrive at the graphene surface in the tangential direction to minimize knock-on collisions.

Figure 1(a) compares the Raman spectra of single-layer graphene before and after tangential- and normal-incidence deposition of Ti. Deposition conditions and film thicknesses are given in the caption. From the much broadened and diminished G and 2D peaks of graphene under normal-incidence sputtering (lower panel in Fig. 1(a)), we see that the graphene lattice has been severely disrupted. The $I_D/I_G$ ratio (~384%) suggests that the average domain size $L_a$ of the resulting graphene is only ~5.0 nm. In contrast, keeping pressure, discharge power, and substrate-target distance constant, the grazing-angle configuration dramatically suppresses impact and preserves the graphene characteristic G and 2D peaks (upper panel in Fig. 1(a)). The $I_D/I_G$ ratio is reduced by a factor of 5 to ~74%, indicating that the graphene domain size has been improved to $L_a \sim 26$ nm. As illustrated in Fig. 1(b), fine-tuning the sputtering parameters and magnetron configurations can further reduce sputtering damages. With this technique, we demonstrate a suppression of $I_D/I_G$ to ≤ 5% ($L_a \geq 385$ nm) with deposition at elevated substrate temperature, visibly below the degree of disorder caused by the X-ray radiation emitted during e-beam evaporation[13] ($I_D/I_G \sim 14\%$, $L_a \sim 137$ nm) under similar deposition rate and time.

To understand why grazing-angle configuration reduces graphene disorder in sputtering, below we describe the experiments that help identifying the cause of graphene damages. The main energy influxes impinging on graphene during sputtering include high-energy ions and electrons, high-energy neutrals (the sputtered target species and sputtering gas species), and radiation. In DC magnetron sputtering on unbiased substrates, high-energy ions of the sputtering gas and the sputtered target species can be first ruled out, as they are strongly confined to the cathode region (target) by the electrostatic field and seldom reach the substrate. As discussed in the Supplementary Material, we can further rule out the impact of plasma radiations and the escaped high-energy electrons.[14] Therefore by elimination, the remaining energetic *neutral* particles dominate the substrate bombardment.[15] Two types of energetic neutrals are created in sputtering. One is the sputtered target species, the kinetic energy of which usually lies below 10 eV.[7] The other is the fast neutrals of the sputtering gas species, say Ar, formed either by recombination of fast Ar+ ions with the secondary electrons near the negatively charged target[16,17] or by collisions of the accelerating Ar+ with the neutral Ar atoms.[18] Because of charge neutrality, these energetic particles escape electrostatic confinement, reflect off the target with a large kinetic energy up to a few hundred eV (the cathode discharge potential), and damage graphene when landing on it.

One can distinguish between the effects of the two energetic neutral species by measuring the changes in Raman $I_D/I_G$ with sputtering gas (Ar) pressure. Figure 2 plots the results of the pressure-dependence study. First, we notice a ~60% decrease in the deposition rate when Ar pressure increases from 3 mTorr to 10 mTorr, indicating substantial scattering of the sputtered target species with background Ar gas. Since increasing Ar pressure thermalizes the sputtered target particles and reduces their kinetic energy appreciably,[19] it would lower the $I_D/I_G$ ratio if the target species is what damages graphene. This, nonetheless, is opposite to the experimental observations in Fig. 2. At constant discharge power density (~2.95 W/cm$^2$ on the target) and film thickness (3 nm – 3.3 nm), $I_D/I_G$ of graphene after Ti deposition at 10 mTorr enhances by ~60% rather than diminish-ing when compared to the 3 mTorr case. This empirically rules out sputtered target species as the cause of damage. Instead we should consider the impact of the energetic Ar neutrals on graphene. Because of their large kinetic energy ($\geq$ 100 eV in our experiments), the fast Ar neutrals have a much smaller scattering cross-section,[16,19,20] which means they undergo substantially fewer collisions with the plasma, whereas the sputtered target neutrals get strongly scattered. As a result, the ratio of the incoming Ar neutral flux to the sputter-deposited target species flux increases with pressure.[16] The higher the Ar pressure is, the larger the fast-neutral bombardment rate on graphene is (relative to the deposition rate). This explains why graphene damage is much more severe during high-pressure deposition as shown in Figure 2.

The empirical identification of highly damaging sputtering gas neutrals is also supported by theoretical findings. According to atomistic simulations of Ar irradiation, the minimum kinetic energy, $KE_{min}$, for an Ar atom or ion to displace one C atom out of graphene is ~33 eV (for particle bombardment at normal incidence).[21,22] In comparison, the energy of the sputtered target species ranges from a few eV up to at most 20 eV.[7] Collisions with the background Ar gas at a few mTorr further bring down the energy.[19] Therefore, energy transfer from the deposited target species to the C atoms remains well below the graphene displacement threshold ($E_d$),[23] too weak to perturb its structure. On the contrary, the fast Ar neutrals travel between the target and substrate without collisions[25] with energy up to hundreds of eV, well exceeding $KE_{min}$. Therefore, these fast sputtering gas neutrals can easily create defects in graphene when coming in at or near normal-incidence, as previously shown in the lower panel of Fig. 1(a).

If we nonetheless place the substrate in the grazing-incidence configuration, the probability of graphene seeing a line of sight of the quasi-ballistic, highly energetic Ar neutrals[25] becomes much reduced. At tangential incidence, the effect of impact reduction is especially pronounced. Moreover, the $E_d$ of graphene[23] is highly anisotropic – ejecting C atoms along the in-plane direction requires much higher energy ($E_d$ ranges from 43 eV to > 700 eV) than along the normal direction ($E_d$ ~ 23 eV),[22,26] which means that the minimum energy required for an Ar atom to displace a C atom rises to hundreds of eV when coming in near the tangential direction.[22] Therefore, even if graphene is exposed to some fast Ar neutrals due to the spread in trajectory during grazing-angle sputtering, a significant fraction of these particles would not have enough kinetic energy to overcome the high in-plane barrier $E_d$, further reducing the probability of graphene damage.

In Figure 3, we compile the dependence of $I_D/I_G$ on the product of deposition time and discharge power density for various thin films deposited under tangential incidence in both systems. This plot demonstrates that regardless of materials, the degree of graphene damage measured by $I_D/I_G$ increases roughly linearly with the power-density-deposition-time product. The higher the discharge power, the higher the incident energy of the fast neutrals is, and therefore the more severe the damage is. Similarly the longer the deposition time, the longer graphene is exposed to the fast neutrals, and again the more extensive the damage is. Hence, to minimize graphene damage for a given film thickness and target material, one has to maximize the deposition rate while keeping the incidence energy low by carefully tuning the pressure and discharge voltage, such that the exposure to high-energy gas neutrals relative to the deposited target species is minimized.[27] Specifi-cally, at a fixed discharge power density, one should deposit at the lowest possible sputtering gas pressure to best preserve graphene structure, as shown in Fig. 2.

We note that moving from the small magnetron guns holding 1" targets to the 2" magnetron guns enables us to sputter at lower pressure (2 mTorr) with much higher rates ($\geq 1$ nm/min). This is because larger-diameter magnetron guns ensure a better balanced magnetic field distribution, which generates stronger confinement of primary electrons and lowers the working gas pressure. Lower sputtering gas pressure strongly reduces the incoming flux of fast neutrals relative to the deposited target species. Consequently, we observe a significant suppression of the residual exposure to high energy Ar neutrals, yielding samples of minimal damage in graphene as shown in Fig. 1(c) and the inset of Fig. 3

We also note that, contrary to our intuition, in 4 mTorr Ar the sputter deposition rate at tangential-incidence remains as high as ~85% of the deposition rate at normal-incidence. The high throughput of the technique is highly desirable. The relatively small decrease is consistent with deposition in the diffusive regime, where the relatively low-energy sputtered target species undergo numerous scattering events (with a mean free path of the order of ~2 cm at ~4 mTorr[18]) and thus are less sensitive to the orientation of the substrate. When sputtering in 2 mTorr Ar, the mean free path goes up to ~4 cm, which is appreciable compared to the substrate-target distance (20 – 23 cm). Consequently, the directionality of the sputtering rate also becomes noticeable. Nevertheless, the reduction in graphene damage significantly outweighs the reduction in deposition rate when the sputtering incident angle approaches 90°. As shown in Supplementary Figure 1, the $I_D/I_G$ ratio decreases by 10-fold when the incident angle changes from 45° to 90° while the deposition rate decreases only by a factor of 2.5 to $\geq 1$ nm/min for Al at 90° tangential incidence.

To evaluate the feasibility of using grazing-angle sputtering for dielectric growth, we studied the film morphology of in-situ oxidized 1 nm Al films deposited at tangential incidence on graphene.[14] Achieving film uniformity is as crucial as minimizing damages when growing thin tunnel barriers or gate oxides. We compared $AlO_x$ on graphene grown at room temperature and at elevated temperature ($T$ = 200 °C). In room-temperature growth, scanning electron microscopy (SEM) and atomic force micrographs (AFM) revealed frequent cracks in $AlO_x$ on graphene with an r.m.s. roughness of $\geq$ 3.18±0.30 Å, indicating that the dielectric coverage can be compromised locally (Supplementary Figure 2(a) and Fig. 4(a)). However, film morphology was dramatically improved in elevated-temperature deposition of Al at $T$ = 200 °C, followed by in-situ oxidation (also at 200 °C) and a gradual cool-down. Moreover, graphene disorder measured by $I_D/I_G$ was further suppressed from ~ 15% to $\leq$ 5% (Fig. 3 inset), presumably a result of the graphene self-healing effect at elevated temperature.[28] We thus observed homogeneous, fully-covered $AlO_x$ films with minute graphene damage and an improved r.m.s. roughness of ~2.46±0.15 Å (Fig. 4(b)), about half the lattice constant of aluminum oxide (a = 4.785Å). These findings demonstrate the potential of using high-temperature grazing-angle sputtering in developing uniform gate dielectrics, tunnel barriers, and possibly also multilayer thin films for graphene device applications.

Recently, flipping-configuration sputtering on graphene was reported to mimic thermal evaporation and eliminate graphene disorder.[10] By facing the substrate 180° away from the target during sputtering at high pressure ($P$ = 20 mTorr), this process creates negligible damage in graphene. However, due to the substrate-target configuration and severe scattering, the deposition rate dropped more than an order of magnitude to 2.5 nm *per* *hour* for Al, normally a high-yield metal.[10] This is in contrast to the 60 nm – 100 nm *per hour* rate in our tangential-incidence configuration. Such low deposition rates render flipping sputtering undesirable for industrial-scale production. In comparison, our grazing-angle sputtering method can be easily scaled up by adopting large-diameter sputter guns and substrate rotations for large-area, high-throughput deposition. Lastly, the nucleation and film growth mode in grazing-angle sputtering better resembles the energetics of conventional sputtering. Therefore, we expect it to produce better controlled magnetic multilayers and tunnel barrier oxides than thermally evaporated deposition methods.

**Acknowledgements:**

The authors would like to thank S. M. Rossnagel, Jani Kotakoski, Hsin-Ying Chiu, Li Yang, K. F. Mak, P. Avouris, and V. Perebeinos for illuminating discussions. We are grateful for the technical support of S. L. Brown, J. R. Rozen, T. Ruiz, and B. A. Ek on modifying the high-vacuum sputtering system, the assistance of S.-J. Han with e-beam evaporation, and the supply of CVD graphene from Satoshi Oida for the incident-angle-dependence studies. E. A. Casu acknowledges the fellowship support of his internship at IBM from the Politecnico di Torino.

fore, the target neutrals get scattered multiple times and thermalized before reaching the substrate.

27. Ideally, one would like to keep the discharge voltage below the displacement threshold $E_d$ of graphene to eliminate C atom ejection at impact. Nevertheless, this voltage threshold is generally too small to maintain stable plasma at a few mTorr. We could increase the sputtering pressure substantially to keep the plasma going, but the resulting deposition rate would be too low for any practical application. Therefore, we have to strike a balance between the discharge voltage and the sputtering pressure. From experience, the sweet spot lies within 2 – 4 mTorr and 100 – 400V for our systems, and the exact values depend on the material type.

**List of Figures:**

**Figure 1 (a)** Raman spectra of monolayer graphene (SLG) before (blue) and after (red) tangential- (upper curves) and normal-incidence (lower curves) sputtering of Ti in 4 mTorr Ar with a discharge power density of ~5.12 W/cm$^2$ on a 1" diameter target. The tangential-incidence data have been shifted up by 600 units for clarity. The film deposition rate is ~0.17 nm/min at tangential incidence and ~0.2 nm/min at normal incidence, and the film thickness is ~5 nm and ~3.6 nm respectively. **Inset:** Schematics of the two sputtering configurations. **(b)** Raman spectra of SLG after tangential-incidence sputtering of 1 nm Al and e-beam evaporation of 1 nm Al. Sputtering was done at 200 °C at the substrate in 2 mTorr Ar with ~2.47 W/cm$^2$ of discharge power density on a 2" target, with a deposition rate of ~1 nm/min, comparable to the E-beam evaporation rate of ~0.6-1.2 nm/min.

**Figure 2.** Raman spectra of single-layer graphene (SLG) after tangential-incidence sputtering of Ti in 3 mTorr (black) and 10 mTorr Ar (red, shifted up by 500 units for clarity). Both are deposited under the discharge power density of ~2.95 W/cm$^2$ on a 1" target. The thickness of the films is 3 nm – 3.3 nm.

**Figure 3.** Dependence of $I_D/I_G$ ratio with discharge-power-density deposition-time product for films deposited by tangential-incidence sputtering. Films grown on graphene in the magnetron sputtering system with 1" targets are labeled in open black diamonds. Films grown in the system with 2" targets are shown in red crosses. The sample-target distance is fixed at ~23 cm, and film thickness is kept below 5 nm. **Inset:** Raman spectra of sputtered, in-situ oxidized 1 nm Al thin films on graphene grown at 200 °C with (upper curve) and without (lower curve) the Ti adhesion layer.

**Figure 4.** AFM micrographs (1 μm × 1 μm) of the in-situ oxidized 1nm Al film on graphene deposited by grazing-angle sputtering at **(a)** room temperature (RT) and at **(b)** 200 °C (HT).

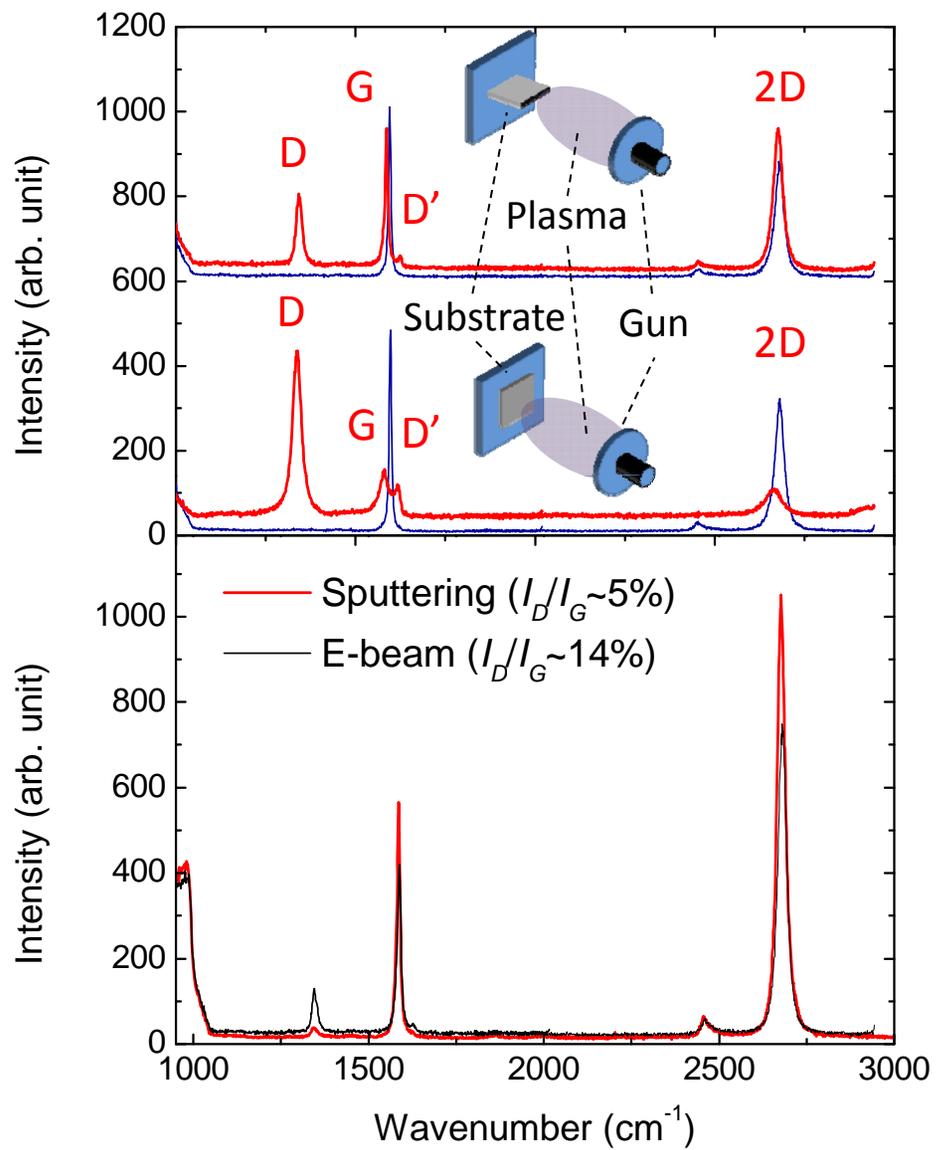

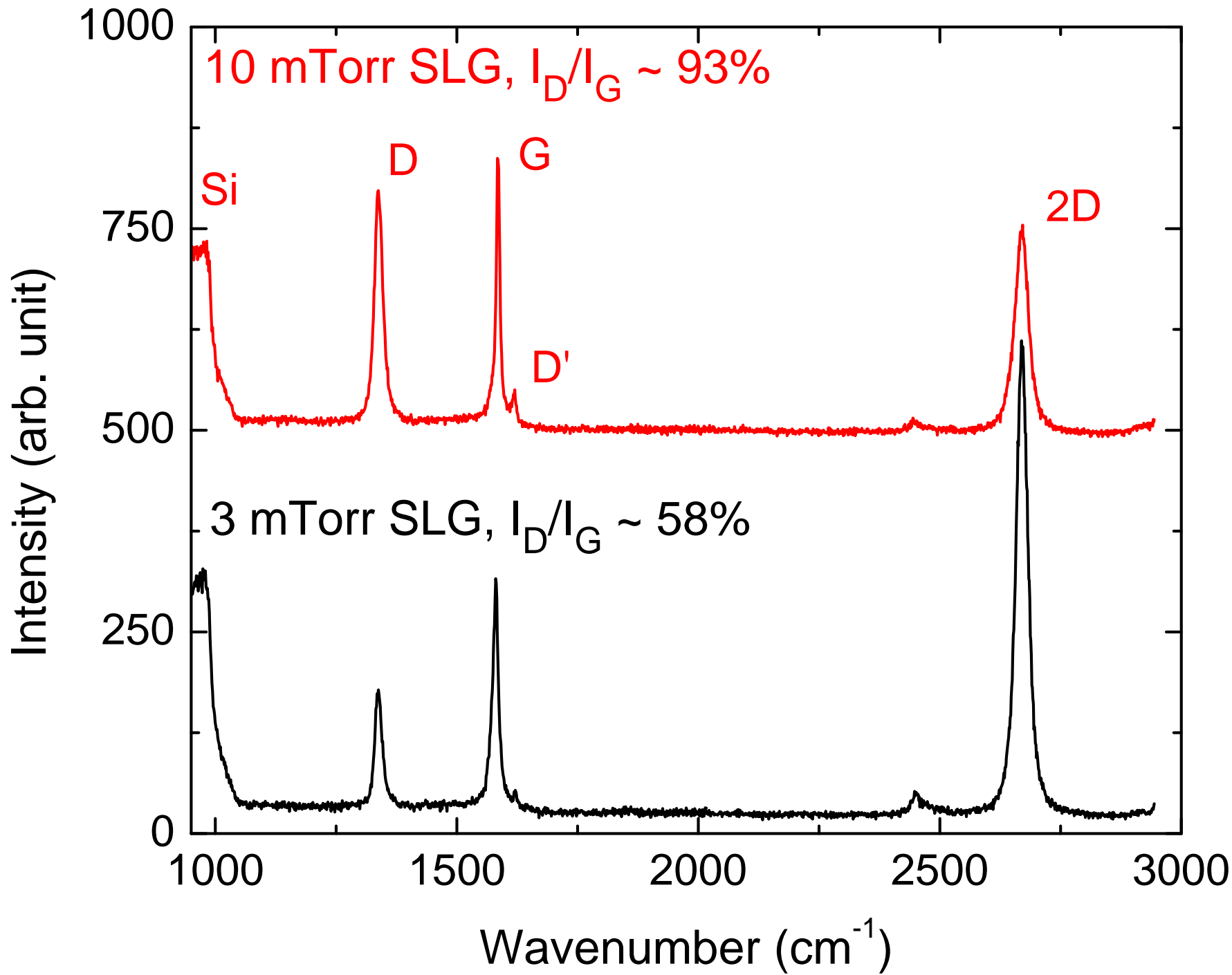

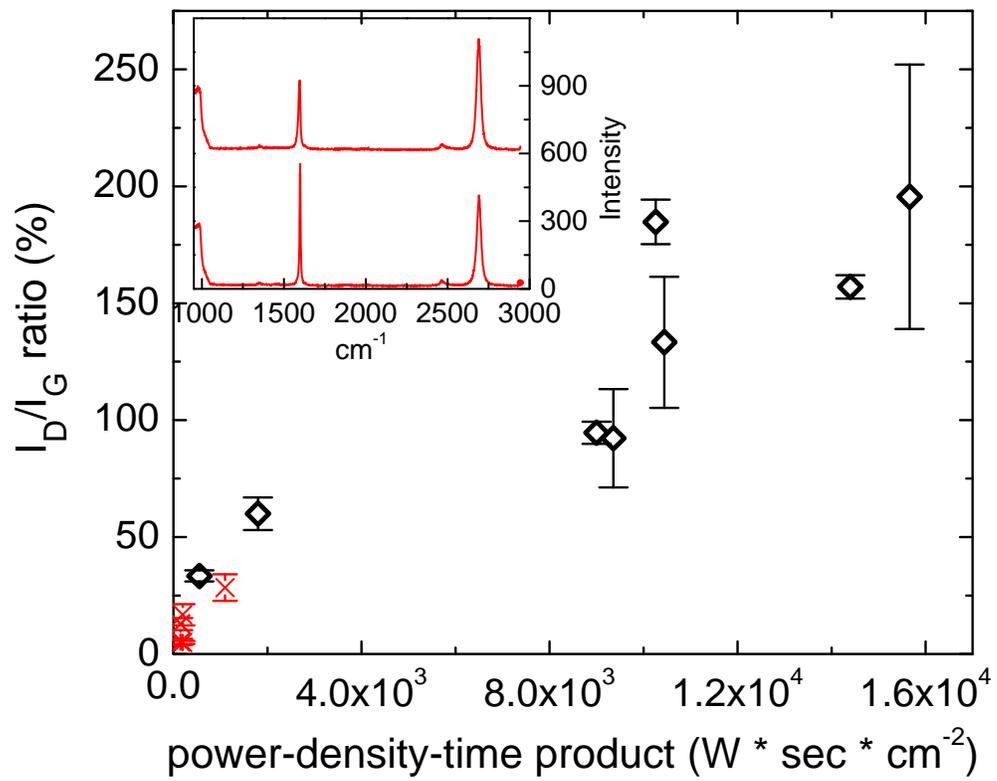

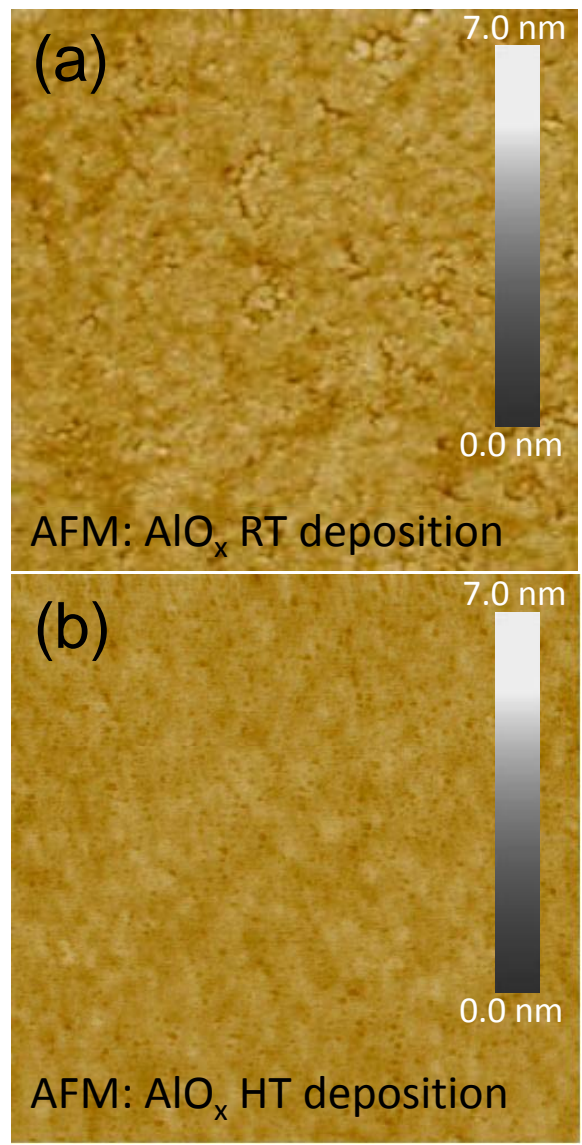

# Low-Damage High-Throughput Grazing-Angle Sputter Deposition on Graphene

C.-T. Chen[1,1], E. A. Casu[1,2,2], M. Gajek[1] and S. Raoux[1]

Supplementary Material

**Supplementary Text – Impact of Plasma Radiation on Graphene:**

We can rule out plasma radiation induced graphene damage for the following reasons. Since photons are massless and transfer negligible kinetic energy, plasma radiation can only create damages when they induce photochemical reactions or generate local heating above the thermal stability threshold in graphene. For photochemical reactions to take place, photon energy has to reach ~12.5 eV so that σ–bond electrons can be promoted to π*–bond.[1] However, the UV photons emitted by Ar plasma and Kr plasma are only 11.6 – 11.8 eV and 10 – 10.6 eV respectively.[2] Therefore, plasma radiation is *not* energetic enough to break the C–C $sp^2$ bonds. Furthermore, graphene optical absorbance between 8 eV to 12 eV is minute.[1] Consequently, radiation induced bond-breaking and heating is also negligible, especially in the case of on-substrate graphene samples, and therefore *cannot* induce structural damages.

**Supplementary Text – Impact of Electron Bombardment on Graphene:**

Next we examine the impact of electron bombardment on graphene. Previous magnetron sputtering experiments have found that escaping primary and secondary electrons may account for a significant fraction of the energy influx towards the substrate.[3] However, to cause damages such as vacancies[4] or bond rotations[5] in graphene during the knock-on collisions, these electrons must transfer sufficient energy to the C atoms to overcome their displacement threshold $E_d$.[6]

---

[1] Email: cchen3@us.ibm.com
[2] Present address: NANOLAB, École Polytechnique Fédérale de Lausanne, Lausanne 1015, Switzerland

Both density-functional molecular dynamics simulation (DFT-MD) and the tight-binding (DFTB) model calculate $E_d$ to be 22-23 eV in graphene.[7-9] One can then estimate the kinetic energy the incident electrons needs to have for generating such a high energy transfer. Using the elastic Coulomb scattering approximation, the maximum energy transfer to the C atoms, $T_{max}$, is related to the incident electron energy $E_e$ by $T_{max} = \frac{2M_C E_e(E_e+2m_e c^2)}{(M_C+m_e)^2 c^2 + 2M_C E_e}$, where $M_C$ is the mass of C atom and $m_e$ is the electron mass.[7-9] To transfer $T_{max} \sim E_d \sim 23$ eV to C atoms for ejection from graphene, it requires ~113 keV of incident electron energy. (We note that because of the large mass mismatch between electrons and C atoms, the energy transfer by collisions is rather inefficient.) In comparison, the energy of the irradiated secondary electrons in sputtering usually lies below 100 eV[10] and that of the escaping primary electrons below the discharging energy (< 500 eV). Both are orders of magnitude lower than the energy value necessary for vacancy generation.[4-6] Even if the increase of energy transfer due to lattice vibrations[4] is taken into account and the lower-transformation-energy bond-rotational Stone-Wales defect[5,11] is considered, the reduced onset energy for emission, 80 – 100 keV,[4,5] still lies well beyond the energy scale of the irradiated electrons in any sputtering system. Therefore, electron bombardment during sputtering does *not* disrupt the graphene lattice structure.

**Supplementary Text – Sputtering Gas Species, Ar vs. Kr:**

We have also investigated the collision impact imparted by a heavier gas species, Kr. Keeping the discharge current, gas pressure, and deposition time constant, we see a larger damage sustained by graphene in Ti/Kr sputtering than that in Ti/Ar sputtering (see Supplementary Table I below). Contrary to the prediction of binary collisions,[12] using heavier sputtering gas appears to provide little advantage when the target species is light. It is also interesting to note that if we consider only binary collision mechanism as many prior studies have assumed,[12-14] the reflection coefficients of heavy sputtering-gas species bouncing off light target elements (such as Ar and Kr bombardment of Al or Ti) should be exceedingly small, especially with Kr gas. Our observation of noticeable sputtering damages in graphene, which is particularly severe under normal incidence (Fig. 1(a) in the main text), indicates that non-binary multiple-collision events play an important role in the reflection of energetic gas neutrals, consistent with the direct particle-flux measurements reported in Ref. 16 and Ref. 17 in the main text.

**Supplementary Text – Ti Adhesion Layer and Film Morphology:**

As noted in the main text, when growing thin dielectrics for tunnel barrier or gate oxide applications, achieving film uniformity is crucial. In the case of ALD of $Al_2O_3$[15] and molecular beam epitaxy (MBE) deposition of MgO[16], dusting the graphene surface with a seed layer is necessary to ensure uniform coverage. Here we summarize the film morphology study of in-situ oxidized 1 nm Al on graphene with and without 1 monolayer (ML) of Ti adhesion layer deposited by tangential grazing-angle sputtering.

While room-temperature as-grown Al oxidized film showed frequent cracks under SEM and AFM (Supplementary Fig. 2(a) and 2(b)), we found that film coverage did not seem to improve much with the Ti layer, but the r.m.s. roughness of locally flat regions (~1 µm$^2$) was reduced from ~3.18±0.30 Å to ~2.76±0.15 Å. On the other hand, when grown at elevated temperature ($T$ = 200 °C), both films with and without the Ti layer showed complete coverage under SEM. Interestingly, the Al film sputtered and oxidized at elevated-temperature ($T$ = 200 °C) exhibited a larger r.m.s. roughness when the Ti layer was present. Film roughness of ~3.90±0.45 Å was observed in AFM (Supplementary Fig. 2(c)), in comparison to ~2.46±0.15 of the $AlO_x$ film without Ti (Supplementary Fig. 2(d)). This morphology study demonstrates that the growth mechanism in grazing-angle sputtering differs significantly from the low-energy ALD and thermal evaporation MBE, and therefore the film nucleation processes are also different.

**Supplementary Figures:**

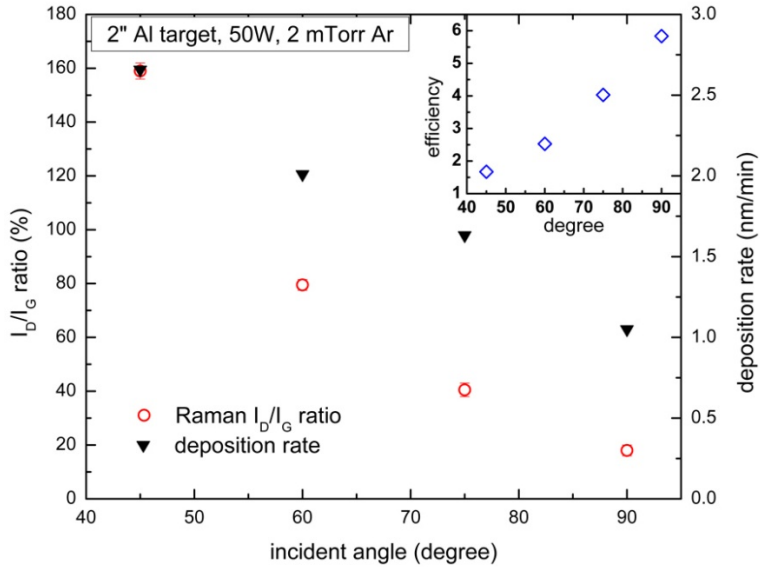

**Supplementary Figure 1.** Angle dependence of the $I_D/I_G$ ratio vs. deposition rate. Data were taken on 1 nm Al films deposited in 2 mTorr Ar with 50 W on a 2" diameter target (~2.47 W/cm$^2$ of discharge power density). The incident angle varies from 45° to 90° in 15° interval. Graphene damage is reduced from $I_D/I_G$ ~160% at 45° to $I_D/I_G$ ~15% at 90°, while the deposition rate decreases from ~2.6 nm/min to ~1 nm/min. **Inset:** Angle dependence of sputtering efficiency. Here we define the efficiency of sputtering (in arbitrary unit) as the deposition rate divided by the $I_D/I_G$ ratio to highlight the gains in damage suppression when sputtering at an angle close to tangential incidence.

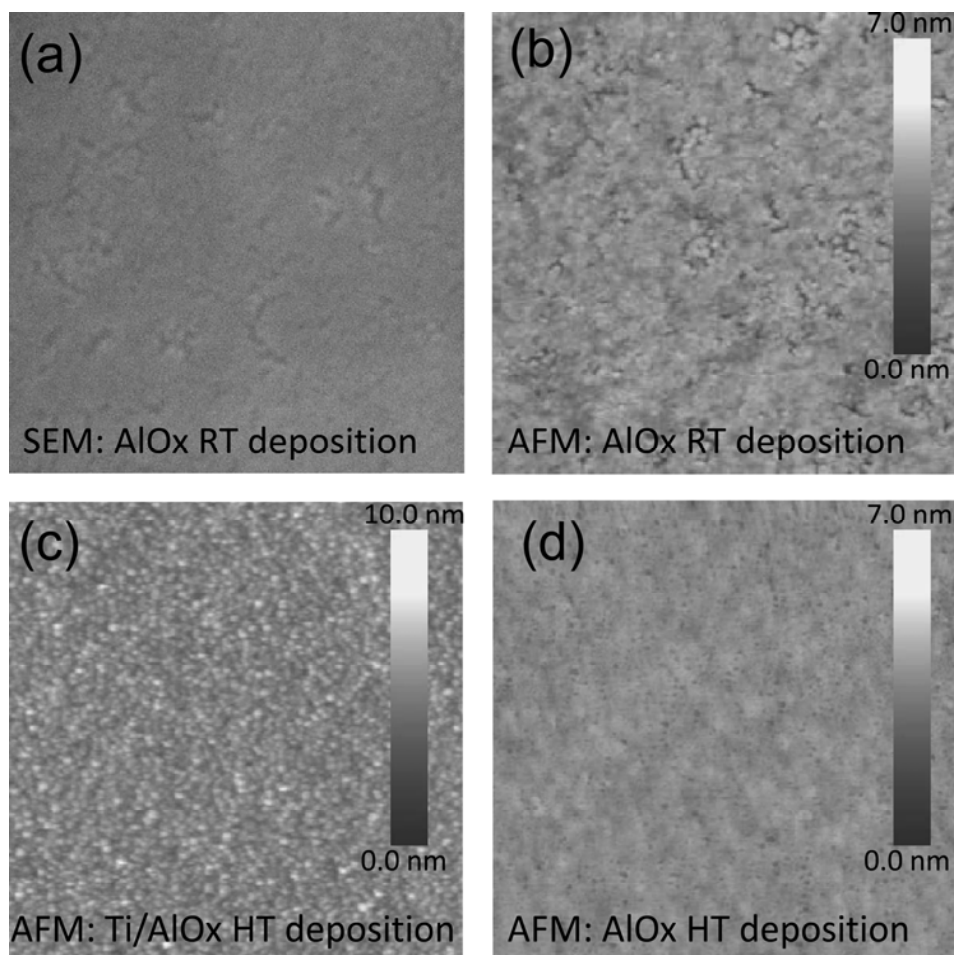

**Supplementary Figure 2.** Morphology of in-situ oxidized AlO$_x$ thin films on graphene deposited by grazing-angle sputtering. **(a)** SEM image (1 µm × 1 µm) of the in-situ oxidized 1nm Al film on graphene deposited at room-temperature (RT). **(b)** AFM micrograph (1 µm × 1 µm) of the in-situ oxidized 1nm Al film on graphene deposited at room temperature (RT). **(c)** AFM micrograph (1 µm × 1 µm) of the in-situ oxidized 1 nm Al film on graphene deposited at 200 °C (HT), with 1monolayer of in-situ oxidized Ti buffer layer. **(d)** AFM micrograph (1 µm × 1 µm) of in-situ oxidized 1nm Al film on graphene deposited at 200 °C (HT).

**Supplementary Table I**. Comparison of sputtering induced graphene damage by Ar and Kr

| Gas Species | Ar | Kr |
|---|---|---|
| Target Species | Ti | Ti |
| Gas Pressure (mTorr) | 4 | 4 |
| Discharge Current (mA) | 40 | 40 |
| Deposition Time (min) | 30 | 30 |
| Film Thickness (nm) | 5 | 3.7 |
| $I_D/I_G$ | 95% ± 4% | 133% ± 23% |